\documentclass[aps, prd, notitlepage, letterpaper, 12pt, amsmath, amssymb, amsfonts, floatfix, nofootinbib, superscriptaddress]{revtex4-1}

\usepackage[utf8]{inputenc}
\usepackage[english]{babel}
\usepackage{microtype}
\usepackage{graphicx} 
\usepackage{dcolumn}
\usepackage{slashed}
\usepackage{lineno}
\usepackage{color}
\usepackage{amsmath}
\usepackage{amssymb}
\usepackage[dvipsnames]{xcolor}
\usepackage{cancel}
\usepackage{framed}
\usepackage{comment}
\usepackage{mathtools}
\usepackage{hyperref}
\usepackage{indentfirst}
\usepackage{multirow}
\usepackage{physics}

\usepackage{ulem}

\definecolor{added}{rgb}{0,0,1}
\definecolor{deleted}{rgb}{1,0,0}

\makeatletter
\newcommand*{\toccontents}{\@starttoc{toc}}
\makeatother

\allowdisplaybreaks

\begin{document}


\title{Flavorful Two Higgs Doublet Models with a Twin}

\author{Wolfgang~Altmannshofer}
\email{waltmann@ucsc.edu}
\affiliation{Department of Physics, University of California Santa Cruz, 1156 High St., Santa Cruz, CA 95064, USA and \\
Santa Cruz Institute for Particle Physics, 1156 High St., Santa Cruz, CA 95064, USA}

\author{Brian~Maddock}
\email{bmaddock@ucsc.edu}
\affiliation{Department of Physics, University of California Santa Cruz, 1156 High St., Santa Cruz, CA 95064, USA and \\
Santa Cruz Institute for Particle Physics, 1156 High St., Santa Cruz, CA 95064, USA}
\begin{abstract}
A two Higgs doublet model with flavorful Yukawa structure, in which the two doublets give mass to the third and the first two generations respectively, is combined with the twin Higgs mechanism to stabilize the Higgs mass against radiative corrections. We consider both a mirror twin and fraternal twin setup. We identify Higgs signal strength measurements and the decay $B_s \to \mu\mu$ as the most important indirect constraints on the parameter space of the model. We explore the collider phenomenology of the model and find that the heavy Higgs in the visible sector can give a sizable number of displaced decays into $b$-jets in regions of parameter space where the SM-like Higgs and the twin Higgs do not provide any striking signatures.
\end{abstract}
\newpage

\maketitle

\section{Introduction} \label{intro}

The absence of clear evidence for new degrees of freedom at the electroweak scale from the Large Hadron Collider (LHC) challenges ``traditional'' solutions to the hierarchy problem that predict new colored degrees of freedom at the TeV scale.
One elegant way to address the hierarchy problem that largely avoids constraints from direct searches at the LHC, is the twin Higgs mechanism~\cite{Chacko:2005pe} and its variations~\cite{Craig:2014aea,Craig:2014roa}. In the twin Higgs model the SM Higgs exists as part of an enlarged, approximately $SU(4)$ symmetric, scalar sector. The symmetry is broken resulting in Higgs doublets both in the visible sector and in an additional ``twin'' sector. The original twin Higgs model prescribed a mirror symmetry, resulting in an exact copy of the SM in the twin sector (see e.g. also~\cite{Barbieri:2005ri,Barbieri:2016zxn}). 
The twin fermions are not charged under the SM gauge symmetries and therefore very hard to search for experimentally.
The original twin Higgs model includes light twin fermions and a massless twin photon. These light degrees of freedom lead to the mirror twin Higgs model having tension with early universe cosmology~\cite{Chacko:2005pe, Craig:2016lyx}. 

Twin Higgs models can be reconciled with cosmological bounds for example in non-standard cosmologies~\cite{Craig:2016lyx,Chacko:2016hvu,Csaki:2017spo, Chacko:2018vss}, or by relaxing the mirror symmetry so that there are no light degrees of freedom in the twin sector. One realization of the second approach is the fraternal twin Higgs (FTH) model~\cite{Craig:2015pha}. In this model the twin sector is constructed with the minimal amount of new physics needed in order to solve the little hierarchy problem in a consistent way. The minimal twin sector required to stabilize the Higgs up to a scale of $O(10)$~TeV contains a twin Higgs doublet, the twin third generation of fermions, and a twin $SU(3)_c \times SU(2)_L$ gauge symmetry. 

Twin Higgs models have been explored extensively in recent years. For example, the collider phenomenology of such models have been studied in~\cite{Craig:2015pha, Curtin:2015fna, Ahmed:2017psb, Kilic:2018sew, Chacko:2019jgi}. Distinct collider signatures arise due to the fact that the SM sector and the mirror sector are only connected through the Higgs portal (see however~\cite{Chacko:2017xpd, Bishara:2018sgl}). 
Twin Higgs models also lead to interesting dark matter phenomenology~\cite{Garcia:2015loa, Garcia:2015toa, Farina:2015uea, Craig:2015xla, Freytsis:2016dgf, Prilepina:2016rlq, Hochberg:2018vdo, Cheng:2018vaj, Terning:2019hgj, Koren:2019iuv}, they can be used to model baryogenesis~\cite{Farina:2016ndq}, and can give rise to exotic astrophysical signatures~\cite{Curtin:2019ngc}.

In the fraternal twin Higgs model the third generation and the first and second generations are inherently treated differently. We wish to motivate the distinction between these generations. We propose that the visible sector is actually realized as a 2 Higgs doublet model (2HDM) with a flavorful Yukawa structure~\cite{Altmannshofer:2016zrn, Altmannshofer:2018bch}. One Higgs doublet is responsible for the mass of the third generation fermions and the other doublet is responsible for the mass of the first and second generations. 
In such a flavorful 2HDM (F2HDM), the mass of the first and second generation of fermions is set by the vacuum expectation value (vev) of the second Higgs that can be considerably smaller than the vev of the first Higgs. Combining the flavorful 2HDM with the twin Higgs mechanisms thus offers the possibility to partially address the hierarchical structure of the quark and charged lepton masses and, at the same time, to stabilize the electroweak scale up to $O(10)$~TeV.

We consider two setups of this ``twinned'' flavorful two Higgs doublet model. In the fist setup, the twin sector is realized in a similar fashion to the mirror twin Higgs model, with a fully mirrored 2HDM structure. In the second setup, we consider a minimal twin sector similar to that of a fraternal twin Higgs model. We show under which conditions these two setups can be mapped onto each other. 

The paper is organized as follows: we briefly summarize twin Higgs models in sec.~\ref{THM}; in sec.~\ref{FT2HDM} we describe the details of the setup of our twin F2HDM and discuss the resulting physical Higgs mass eigenstates and their couplings to both the SM and twin sector particles; in sec.~\ref{Consts} we discuss the bounds on the model from Higgs signal strength measurements and the most important flavor constraint, the $B_s \to \mu \mu$ decay; finally, in sec.~\ref{pheno} we look at the phenomenology of this model, particularly focusing on displaced decays occurring in regions of parameter space that are unique to this setup; we conclude in sec.~\ref{sec:conclusions}. 

\section{Twin Higgs Models}
\label{THM}

The twin Higgs mechanism stabilizes the Higgs mass up to some moderate scale, $\Lambda$, usually considered to be around $10$~TeV. Above this scale some additional new physics is invoked to protect the Higgs mass up to the Planck scale. The largest contributions to the Higgs mass are the 1-loop top quark correction, the 1-loop $SU(2)_L$ correction, and the 2-loop QCD correction. In the twin Higgs model a twin sector exists with new degrees of freedom which cancel these contributions. 
Here we briefly review two versions of the twin sector: the mirror model and the fraternal model. More detailed discussions of these models and the underlying protection mechanism can be found in~\cite{Chacko:2005pe} and~\cite{Craig:2015pha}, respectively. 
 
The twin Higgs mechanism is based on an approximate $SU(4)$ symmetry that is respected by the scalar sector. An $SU(4)$ fundamental scalar $\Phi$ contains two doublets $\phi$ and $\hat \phi$, parameterized as 
  \begin{equation}
  \Phi = \begin{pmatrix}
       \phi \\
       \hat{\phi}
      \end{pmatrix} = 
      \begin{pmatrix}
       \phi^+ \\
       (v + S + i\eta)/\sqrt{2} \\
\hat{\phi}^+ \\
(\hat{v} + \hat{S} + i\hat{\eta})/\sqrt{2} 
      \end{pmatrix},
\end{equation}
with the potential
\begin{eqnarray}
	V(\phi, \hat{\phi}) &=& -\mu^2 |\Phi|^2 + \lambda |\Phi|^4 + \kappa |\phi|^4 + \hat{\kappa} |\hat{\phi}|^4 - \sigma \mu^2 |\phi|^2\,. 
\end{eqnarray}
Besides the SU(4) symmetric mass term $\mu^2$ and the quartic coupling $\lambda$, the potential includes a soft $SU(4)$ breaking term $\sigma$, which allows a misalignment of the SM and twin vevs, $v$ and $\hat v$, and the parameters $\kappa$ and $\hat{\kappa}$ are hard breaking terms, which help to reduce fine tuning~\cite{Ahmed:2017psb}. We identify $\phi$ as the $SU(2)_L$ Higgs doublet in the SM sector and $\hat{\phi}$ is the corresponding doublet in the twin sector.

After symmetry breaking and rotating to the physical mass eigenstates results in two physical scalar bosons that we identify as a SM-like Higgs ($h$) and a twin Higgs ($\hat{h}$) which are mixed states of $S$ and $\hat{S}$. The mixing angle is of order $O(v/\hat v)$. 

The particle content of the twin sector is where the mirror and fraternal realizations of the twin Higgs mechanism differ. We first consider the mirror twin Higgs model where the twin sector is an exact copy of the SM sector containing the same forces, particles, and couplings that the SM does. 

\begin{figure}[tb]
\begin{center} 
\includegraphics[width=0.9\textwidth]{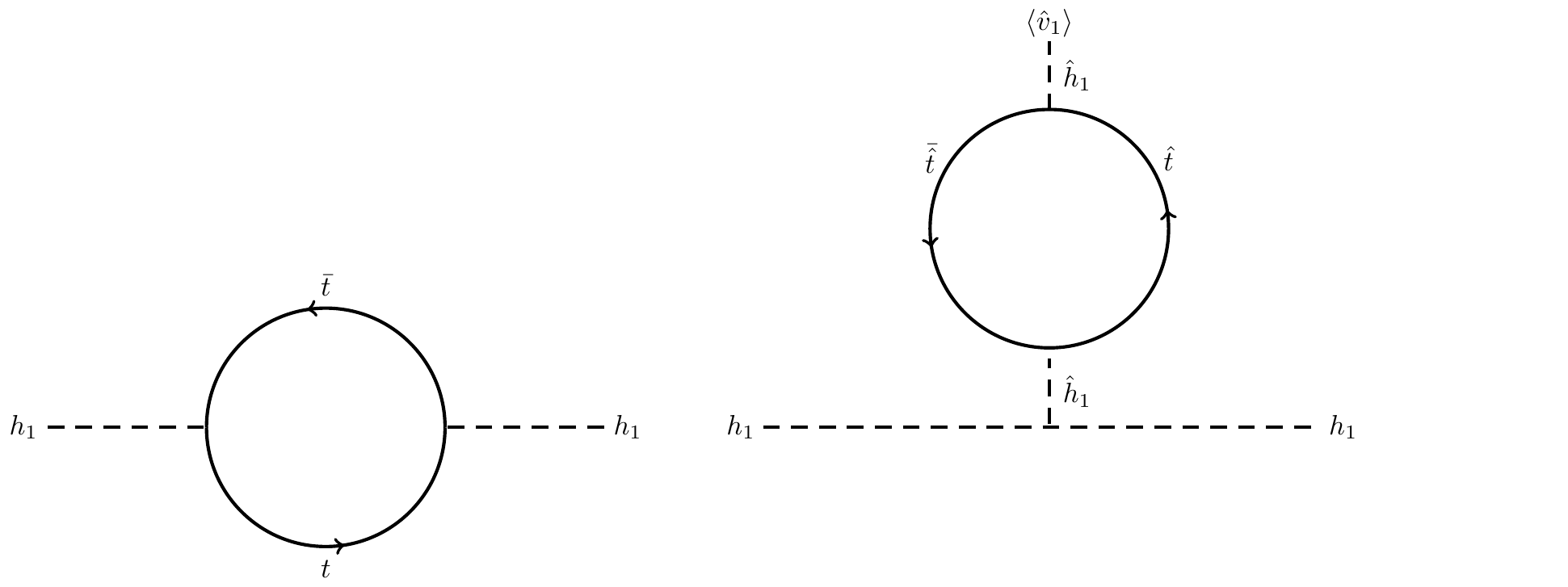}
\caption{Diagrams showing the loop contributions to the Higgs mass from the top quark (left) and from the twin top (right).}
\label{top_loop}
\end{center}
\end{figure}

The Higgs mass receives loop contributions from both fermions and twin fermions, as shown in fig.~\ref{top_loop}. The twin top contribution comes with a relative minus sign as compared to the top contribution causing these two diagrams to cancel. In a similar fashion to the top quarks the twin contributions from the weak gauge bosons and two loop gluon contributions to the Higgs mass are exactly the same as the SM contribution with a relative minus sign. This is the fundamental mechanism that stabilizes the Higgs mass in twin Higgs models. The total correction to the Higgs mass from these loops is~\cite{Craig:2015pha}
 \begin{eqnarray}
 \delta m_h^2 &=&\sum_{f} \frac{N_f \Lambda^2}{4\pi^2}(y_f^2 - \hat{y}^2_{\hat{f}})  +  \frac{9 \Lambda^2}{32\pi^2}(g_2(\Lambda)^2 - \hat{g}_2(\Lambda)^2) + \frac{3y_t^2 \Lambda^2}{4\pi^4}(g_3(\Lambda)^2 - \hat{g}_3(\Lambda)^2), 
 \label{HiggsMass}
 \end{eqnarray}
where $\hat{g}_2$ is the strength of the twin $SU(2)_L$, $\hat{g}_3$ is the strength of the twin $SU(3)$, and $\hat{y}_{\hat{f}}$ are the twin Yukawa couplings. The color factor $N_c = 3$ for quarks and $N_c = 1$ for leptons. In the mirror twin Higgs model the couplings in the twin sector and the visible sector are set to be equal, thus leading to $\delta m_h^2$ above being zero. However, the many light degrees of freedom in the mirror twin sector (in particular the light twin fermions and the massless twin photon) lead to tensions with cosmology. This inspired a minimal version of the twin Higgs model known as the fraternal twin Higgs model. 

The fraternal twin Higgs model adds the minimum new physics in the twin sector necessary to stabilize the Higgs. The particle content in the twin sector consists of a twin top, a twin $SU(2)_L$ and a twin $SU(3)$. In this setup the couplings of these particles are free parameters. From eq.~(\ref{HiggsMass})  we see that to ensure that the Higgs mass is not significantly tuned up to $\Lambda \sim 10$~TeV one requires
 \begin{equation}
\Bigg |\frac{ \hat y_t(\Lambda) - y_t(\Lambda)}{y_t(\Lambda)}\Bigg| \lesssim 0.01, ~~~ \Bigg |\frac{ \hat g_2(\Lambda) - g_2(\Lambda)}{g_2(\Lambda)}\Bigg| \lesssim 0.1,~~~ \Bigg |\frac{ g_3(\Lambda) - \hat g_3(\Lambda)}{g_3(\Lambda)}\Bigg|\lesssim 0.1. 
\label{yConst}
 \end{equation}

In order for the twin $SU(3)$ to be anomaly free there must exist a right handed twin bottom. For the twin $SU(2)$ to be anomaly free one need an $SU(3)$ neutral $SU(2)$ doublet, which contains the left handed twin tau and twin neutrino. In order to make the twin tau and neutrino massive one also introduces a right handed twin tau and twin neutrino. Thus, the minimal particle content of the fraternal twin Higgs model contains a twin Higgs doublet, a full third generation of twin fermions, and twin gauge interactions based on the gauge groups $SU(2)$, and $SU(3)$.

To ensure the twin fermions other than the top do not reintroduce large corrections to the Higgs mass one has to demand that 
\begin{equation}\label{yfConst}
 \hat{y}_{\hat{f}}^2 \lesssim  \frac{4 \pi^2}{3 \Lambda^2} \delta m_h^2 + y_{f}^2 \sim 0.002 \times \left( \frac{10\,\text{TeV}}{\Lambda} \right)^2 \left( \frac{\delta m_h}{125\,\text{GeV}} \right)^2 \,, 
\end{equation}
where in the last step we neglected the small SM Yukawas $y_{f}$. The above criterion translates into $\hat{y}_{\hat{f}}$ being no larger than $\sim 0.05$, with the precise value depending on the maximum acceptable choice for $\delta m_h$.

\section{Twin Two Higgs Doublet Models}
\label{FT2HDM}

Both the mirror and fraternal twin Higgs models successfully stabilize the Higgs mass up to order $\Lambda$. However, the mirror twin Higgs needs additional physics which can reconcile the model with cosmology, while the fraternal twin Higgs model leaves us with no explanation for the lack of the first two generations in the twin sector. Here we describe how the addition of a new source of mass generation in the form of a second Higgs doublet might provide a resolution to these issues. We also show how the mirror and fraternal version of a 2HDM setup can be mapped onto one another. 

\subsection{Mirror Setup}

A well studied setup that provides additional sources of mass generation and distinguishes between the first two generations and the third generation is the flavorful 2HDM \cite{Altmannshofer:2016zrn,Altmannshofer:2018bch}. This model contains a SM-like doublet which primarily provides mass to the third generation fermions, and an additional doublet that primarily provides mass to the first and second generations. We propose a mirror twin Higgs inspired model where both the visible sector and the twin sector are realized as flavorful 2HDMs.

In this realization we have four doublets  $\phi_1, \hat{\phi}_1, \phi_2$ and, $\hat{\phi}_2$, where $\phi_1$ and $\phi_2$ live in the visible sector and $\hat{\phi_1}$ and $\hat{\phi_2}$ live in the twin sector. The fields are arranged into $SU(4)$ multiplets
\begin{eqnarray}
\Phi_1 = \begin{pmatrix}
       \phi_1 \\
       \hat{\phi}_1
      \end{pmatrix}, ~~~~~ \Phi_2 = \begin{pmatrix}
       \phi_2 \\
       \hat{\phi}_2
      \end{pmatrix}.
\end{eqnarray}
We will consider a scenario in which $\phi_1$ and $\hat{\phi}_1$ couple to the third generation particles in the visible and twin sector, respectively, and $\phi_2$ and $\hat{\phi}_2$ couple to the first two generations in the visible and twin sector, respectively. The most generic potential for $\Phi_1$ and $\Phi_2$ looks like
\begin{eqnarray}
V(\Phi_1, \Phi_2)&=& -\mu_1^2 |\Phi_1|^2 - \mu_2^2|\Phi_2|^2 + \lambda_1(\Phi_1^\dagger \Phi_1)^2 + \lambda_2(\Phi_2^\dagger\Phi_2)^2 + \lambda_3 |\Phi_1|^2 |\Phi_2|^2 \nonumber \\
&+& m_{12}^2 \Bigg (\phi_1^\dagger  \phi_2+ h.c.\Bigg ) + \hat{m}_{12}^2 \Bigg ( \hat{\phi}_1^\dagger \hat{\phi}_2 + h.c.\Bigg ) + \lambda_4 |\Phi_1^\dagger \Phi_2|^2 + \frac{\lambda_5}{2}\Bigg ( (\Phi_1^\dagger \Phi_2 )^2 +h.c. \Bigg) \nonumber \\
&+& \kappa_1 (\phi_1^\dagger \phi_1)^2+ \hat{\kappa}_1(\hat{\phi}_1^\dagger \hat{\phi}_1)^2 + \kappa_2(\phi_2^\dagger \phi_2)^2 + \hat{\kappa}_2(\hat{\phi}_2^\dagger \hat{\phi}_2)^2 + \sigma_1 \mu_1^2 (\phi_1^\dagger \phi_1) + \sigma_2 \mu_2^2(\phi_2^\dagger \phi_2)~, 
\label{TF2HDM}
\end{eqnarray}
where the soft breaking terms $\sigma$ and $\hat{\sigma}$ and the hard breaking terms $\kappa_1, \kappa_2$, $\hat{\kappa}_1$, and $\hat{\kappa}_2$ are introduced in analogy to the usual twin Higgs setup. The terms containing $m^2_{12}$ and $\hat{m}^2_{12}$ are mass parameters that mix the doublets $\phi_1$ and $\hat{\phi}_1$ with $\phi_2$ and $\hat{\phi}_2$, respectively. $m^2_{12} \neq \hat{m}^2_{12}$ is another source of soft symmetry breaking.

As shown in \cite{Yu:2016swa} this setup is a self consistent extension of the twin Higgs model and provides the same cancellations as in the traditional twin Higgs setup. However, we now have extra sources of mass generation in the SM sector from $\phi_2$ and the twin sector from $\hat{\phi}_2$. 
Constraints from cosmology (in particular $N_{eff}$) can be avoided by making the twin degrees of freedom sufficiently heavy, i.e. heavier than $O(1\,\text{GeV})$. 

The first two generations in the visible and twin sector have masses 
\begin{equation}
   m_{f_{1,2}}  = \frac{y_{f_{1,2}} v_2}{\sqrt{2}} ~,~~~ m_{\hat{f}_{1,2}}  = \frac{y_{f_{1,2}} \hat{v}_2}{\sqrt{2}} ~. 
\end{equation}
Due to the mirror symmetry the only way to make the first two generations of twin fermions heavy is to make the vacuum expectation value $\hat{v}_2$ much larger than $v_2$ (see~\cite{Batell:2019ptb} for a different mechanism to raise the masses of the fermions in the twin sector.) Characteristic values for  $v_1/v_2 \approx 10$ (motivated to explain the hierarchy between the third and the second generation of SM fermions) and requiring that the lightest mirror particles to be at least $O(1\,\text{GeV})$, leads to $\hat{v}_2 \approx 10$ TeV. We thus envision the following set of vevs
\begin{equation}
  v_2 \sim O(10\,\text{GeV}) ~,~~~ v_1 \sim O(100\,\text{GeV}) ~,~~~ \hat v_1 \sim O(1\,\text{TeV}) ~,~~~ \hat v_2 \sim O(10\,\text{TeV}) ~.
\end{equation}
We can approximate the amount of fine tuning needed to put the vevs in this hierarchical structure as~\cite{Craig:2015pha,Beauchesne:2015lva}
\begin{equation}
v_1-\text{Tuning} \sim \frac{2 v_1^2}{\hat{v}_1^2} ~,~~~ v_2-\text{Tuning} \sim \frac{2 v_2^2}{\hat{v}_2^2} ~. 
\end{equation}
This means the tuning of $v_1$ vs. $\hat v_1$ is order percent level, but the tuning of $v_2$ vs. $\hat v_2$ is substantial, of order $10^{-6}$. 

In addition to the fermions, also the twin photon needs to be sufficiently heavy to avoid cosmological bounds. Two options to do this are: breaking electromagnetism in the mirror sector, or simply removing the $U(1)$ hypercharge in the twin sector. In both cases the mirror symmetry of the model is weakened. In the following we will follow the scenario where there is no twin $U(1)$ hypercharge. 

The setup we have described so far leads to a large number of $O(1\,\text{GeV})$ particles in the twin sector resulting in a complicated, yet rich set of dynamics. We leave a detailed discussion of this scenario to future work. Instead, we focus on a simplified setup which takes the twin Yukawas as free parameters in order to make the first and second generation twin fermions sufficiently heavy to be irrelevant for the Higgs phenomenology that we will discuss below. 

By taking $\hat{y}_{f}$ to be free parameters (up to the bound imposed by eq.~(\ref{yfConst})) we can push the masses of the twin first and second generation particles to $O(1\,\text{TeV})$, for $\hat{v}_2 \sim 10$ TeV. In such a setup, the low energy phenomenology will be determined by the twin third generation, the twin $SU(3)$ and $\hat{\phi}_1$, while all the other twin states are effectively decoupled.

\subsection{Fraternal Setup}

Another approach to a twin flavorful 2HDM is to construct a model inspired by the fraternal twin Higgs.
Starting from a flavorful 2HDM with doublets $\phi_1$ and $\phi_2$, we add a third doublet $\hat{\phi}_1$, with $\phi_1$ and $\hat{\phi}_1$ being part of an approximate $SU(4)$ symmetry. The doublets $\phi_1$ and $\hat{\phi}_1$ are responsible for the mass generations of the third generation particles in their respective sectors. The visible sector doublet $\phi_2$ provides mass to the first and second generation fermions in the visible sector, which have no counterparts in the mirror sector. The twin sector consists of a Higgs doublet, a twin $SU(2)$, a twin $SU(3)$, and the third generation of twin fermions. 

\subsection{Twin F2HDM}

The two approaches mentioned above both result in the same particle content and forces at low scales.
Regardless of the high scale setup we will refer to the low energy simplified model as the twin F2HDM.\footnote{One difference between the two discussed setups (mirror and fraternal twin sector) is that the mass of the twin weak gauge bosons will be set by a combination of $\hat{v}_1$ and $\hat{v}_2$ in the mirror setup, but only set by $\hat{v}_1$ in the fraternal setup. Generically, because $\hat{v}_2 \gg \hat{v}_1$ then twin weak gauge bosons in the mirror setup will be much heavier than in the fraternal setup. However, in both cases the twin weak gauge bosons will be $O(1 \, \text{TeV})$ or heavier, leaving no noticeable difference in the low energy phenomenology we will discuss in the remainder of this paper.}

The potential for the twin F2HDM can be derived from eq.~(\ref{TF2HDM}) with $\hat{\phi}_2$ integrated out. This leaves an effective three Higgs doublet potential for the fields
\begin{equation}
  \Phi_1 = \begin{pmatrix}
       \phi_1 \\
       \hat{\phi}_1
      \end{pmatrix} = 
      \begin{pmatrix}
       \phi_1^+ \\
       (v_1 + S_1 + i\eta_1)/\sqrt{2} \\
\hat{\phi}^+_1 \\
(\hat{v}_1 + \hat{S}_1 + i\hat{\eta}_1)/\sqrt{2} 
      \end{pmatrix}~,~~~ \phi_2 = 
      \begin{pmatrix}
       \phi_2^+ \\
       (v_2 + S_2 + i\eta_2)/\sqrt{2} \end{pmatrix} ~,
\end{equation}
\begin{eqnarray}
V(\Phi_1, \phi_2)&=& -\mu_1^2 |\Phi_1|^2 - \mu_2^2|\phi_2|^2 + \lambda_1(\Phi_1^\dagger \Phi_1)^2 + \lambda_2(\phi_2^\dagger\phi_2)^2 + \lambda_3 |\Phi_1|^2 |\phi_2|^2 \nonumber \\
&+& m_{12}^2 \Bigg (\phi_1^\dagger \phi_2 + h.c.\Bigg ) + \lambda_4 |\phi_1^\dagger \phi_2|^2 + \frac{\lambda_5}{2}\Bigg ( (\phi_1^\dagger \phi_2 )^2 +h.c. \Bigg) \nonumber \\
&+& \sigma \mu_1^2 \phi_1^\dagger \phi_1 + \kappa_1 (\phi_1^\dagger \phi_1)^2 + \kappa_2 (\phi_2^\dagger \phi_2)^2.
\label{FTF2HDM}
\end{eqnarray}
After electroweak symmetry breaking we are left with 6 massive modes: three scalar Higgs bosons, two charged Higgs bosons, and one pseudoscalar Higgs boson. The three scalars $S_1$, $\hat S_1$, $S_2$ are related to the mass basis counterparts $h_1$, $\hat{h}_1$, and $h_2$ (identified as the SM-like, twin, and heavy Higgs) by
\begin{subequations}
\begin{eqnarray}
S_1 &=& c_{\alpha_1} c_{\alpha_2} h_1  +  \Big( c_{\alpha_1} s_{\alpha_2} c_{\alpha_3} + s_{\alpha_1} s_{\alpha_3} \Big) h_2  + \Big( s_{\alpha_1} c_{\alpha_3} - c_{\alpha_1} s_{\alpha_2} s_{\alpha_3} \Big) \hat h_1 ~,\\
\hat{S}_1&=& -s_{\alpha_1} c_{\alpha_2} h_1  +  \Big( c_{\alpha_1} s_{\alpha_3} - s_{\alpha_1} s_{\alpha_2} c_{\alpha_3} \Big) h_2  + \Big( c_{\alpha_1} c_{\alpha_3} + s_{\alpha_1} s_{\alpha_2} s_{\alpha_3} \Big) \hat h_1 ~,\\
S_2 &=& - s_{\alpha_2} h_1 + c_{\alpha_2} c_{\alpha_3} h_2 - c_{\alpha_2} s_{\alpha_3} \hat h_1~.
\end{eqnarray}
\end{subequations}
where the three mixing angles ($s_{\alpha_i} = \sin(\alpha_i)$, $c_{\alpha_i} = \cos(\alpha_i)$) are approximately given by
\begin{eqnarray} \label{eq:angles}
\sin(\alpha_1) &\simeq& \frac{v_1 \lambda_1}{\hat{v}_1(\kappa_2 + \lambda_1)} ~,~~~ \sin(\alpha_2) \simeq -\frac{v_2}{v_1} ~,~~~ \sin(\alpha_3) \simeq -\frac{v_2 \lambda_3}{2 \hat{v}_1 (\kappa_2 + \lambda_1)} ~. 
\end{eqnarray}
The three Higgs boson masses are approximately 
\begin{eqnarray}
 m_{h_1}^2 &\simeq&  2 v_1^2\Bigg(\kappa_1+ \frac{\lambda_1\kappa_2}{\kappa_2 +\lambda_1}\Bigg) ~,~~~ m_{h_2}^2 \simeq \frac{m_{12}^2 v_1}{v_2} ~,~~~ m_{\hat{h}_1}^2 \simeq 2 \hat{v}_1^2 (\lambda_1+\kappa_2) ~.
 \label{higgsMass}
 \end{eqnarray}
The SM-like Higgs mass can be set by fixing $\kappa_1$, $\kappa_2$, and $\lambda_1$. The heavy Higgs mass is primarily set by the parameter $m^2_{12}$, and the twin Higgs mass is primarily set by $\hat{v}_1$, both of which can be taken as free parameters. 

The most generic Yukawa Lagrangian can be written as
\begin{eqnarray}
 -{\cal L}_\text{twin-F2HDM} &\supset&\Bigg \{\sum_{i,j} \left( \lambda^u_{1,ij} (\bar q_i u_j) \tilde\phi_1 + \lambda^d_{1,ij} (\bar q_i d_j) \phi_1 + \lambda^e_{1,ij} (\bar\ell_i e_j) \phi_1 \right) \nonumber \\
 && + \sum_{i,j} \left( \lambda^{u}_{2, ij} (\bar q_i u_j) \tilde \phi_2 + \lambda^{d}_{2, ij} (\bar q_i d_j) \phi_2 + \lambda^{e}_{2, ij} (\bar\ell_i e_j) \phi_2 \right) ~\nonumber \\
 && +\left(\hat{y}_{\hat{t}} (\bar{\hat{q}} \hat{t}) \tilde{\hat{\phi}}_1 +\hat{y}_{\hat{b}} (\bar{\hat{q}} \hat{b})\hat{\phi}_1 + \hat{y}_{\hat{\nu}} (\bar{\hat{\ell}} \hat{\nu}) \tilde{\hat{\phi}}_1+ \hat{y}_{\hat{\tau}}(\bar{\hat{\ell}} \hat\tau) \hat{\phi}_1 \right) +\text{h.c}. \Bigg\}~. 
  \label{eq:Lagrangian}
\end{eqnarray}
The Yukawa matrices in the SM sector $\lambda^f_i$ are determined by the flavor structure imposed on $\phi_1$ and $\phi_2$ in the flavorful setup~\cite{Altmannshofer:2016zrn, Altmannshofer:2018bch}. The couplings $\lambda^u_1$, $\lambda^d_1$, and $\lambda^e_1$ are rank one matrices, providing mass only to the third generation, while $\lambda^u_2$, $\lambda^d_2$, and $\lambda^e_2$ have full rank and provide mass for the remaining fermions as well as CKM mixing in the quark sector. We find that the couplings of the Higgs bosons to the up-type quarks in the fermion mass eigenstate basis are given by
\begin{subequations}
\begin{eqnarray}
Y^{h_1}_{u_i u_j}  &=& \delta_{ij} \frac{m_{u_i}}{v~s_\beta}(c_{\alpha_1} c_{\alpha_2}) - \frac{m_{u_i u_j}}{v~s_\beta c_\beta }(c_\beta c_{\alpha_1} c_{\alpha_2} + s_\beta s_{\alpha_2}) ~,\\
Y^{h_2}_{u_i u_j} &=&\delta_{ij} \frac{m_{u_i}}{v~s_\beta} (s_{\alpha_3} s_{\alpha_1} + c_{\alpha_1}s_{\alpha_2}c_{\alpha_3})-\frac{m_{u_i u_j}}{v~s_\beta c_\beta}(- c_{\alpha_2}c_{\alpha_3}s_\beta + c_{\alpha_1}c_{\alpha_3}s_{\alpha_2} c_\beta + s_{\alpha_1}s_{\alpha_3} c_\beta) ~,\\
Y^{\hat{h}_1}_{u_i u_j} &=&  \delta_{ij}\frac{m_{u_i}}{v~s_\beta} (c_{\alpha_3} s_{\alpha_1} - c_{\alpha_1}s_{\alpha_2}s_{\alpha_3})+\frac{m_{u_i u_j}}{v~s_\beta c_\beta}(c_{\alpha_1}s_{\alpha_2}s_{\alpha_3}c_\beta - c_{\alpha_3}s_{\alpha_1}c_\beta - c_{\alpha_2}s_{\alpha_3} s_\beta) ~,
\end{eqnarray}
\label{fermionCouplings}
\end{subequations}
where $v = \sqrt{v_1^2 + v_2^2} = 246$~GeV and $s_\beta = \sin\beta$, $c_\beta = \cos\beta$ with $\tan\beta = v_1/v_2$. The mass parameters $m_{u_i u_j}$ are given by the Yukawa couplings $\lambda^u_2$ in the fermion mass eigenstate basis. For the flavor indices $i$ or $j$ equal to 1, the mass parameters $m_{u_i u_j}$ are of the order of the up quark mass and of the order of the charm quark mass otherwise (see~\cite{Altmannshofer:2018bch} for their explicit expressions). 

The above expressions for the couplings hold analogously for the down-type quarks and leptons. The couplings of the SM fermions to the charged Higgs bosons are the same as in the standard versions of the F2HDMs~\cite{Altmannshofer:2016zrn}. In our setup discussed here, the scalar Higgs bosons (and charged Higgs bosons) couple in addition also to the twin sector fermions as
\begin{subequations}
\begin{eqnarray}
Y^{h_1}_{\hat{f}} &=& -\frac{\hat y_{\hat{f}}}{\sqrt{2}} c_{\alpha_2} s_{\alpha_1} ~, \\
Y^{h_2}_{\hat{f}} &=& -\frac{\hat y_{\hat{f}}}{\sqrt{2}} (c_{\alpha_3}s_{\alpha_1} s_{\alpha_2} - c_{\alpha_1} s_{\alpha_3}) ~,\\
Y^{\hat{h}_1}_{\hat{f}} &=&\frac{\hat y_{\hat{f}}}{\sqrt{2}} (c_{\alpha_1} c_{\alpha_3} + s_{\alpha_1} s_{\alpha_2} s_{\alpha_3}) ~. 
\end{eqnarray}
\label{twinCouplings}
\end{subequations}
Finally, the couplings of the Higgs bosons to the vector bosons ($hWW$ and $hZZ$) are given by the following expressions
\begin{subequations}
\begin{eqnarray}
\frac{Y^{h_1}_{V}}{Y^\text{SM}_{V}} &=& c_{\alpha_1} c_{\alpha_2} s_\beta - s_{\alpha_2} c_\beta ~,\\ 
\frac{Y^{h_2}_{V}}{Y^\text{SM}_{V}} &=&  (c_{\alpha_1}c_{\alpha_3}s_{\alpha_2} + s_{\alpha_1}s_{\alpha_3})s_\beta -(c_{\alpha_3} s_{\alpha_1}s_{\alpha_2} - c_{\alpha_1}s_{\alpha_3}   ) c_\beta ~, \\
\frac{Y^{\hat{h}_1}_{V}}{Y^\text{SM}_{V}} &=& (c_{\alpha_3} s_{\alpha_1} - c_{\alpha_1}s_{\alpha_2} s_{\alpha_3} ) s_\beta + (c_{\alpha_1} c_{\alpha_3} + s_{\alpha_1} s_{\alpha_2} s_{\alpha_3}) c_\beta~.
\end{eqnarray}
\label{vectorCouplings}
\end{subequations}
where $Y^\text{SM}_{V}$ are the corresponding couplings of the Higgs boson in the Standard Model.

\section{Constraints}
\label{Consts}
The introduction of two additional Higgs doublets alters the couplings of the SM-like Higgs boson $h_1$ as shown in eqs.~(\ref{fermionCouplings})-(\ref{vectorCouplings}). The ATLAS and CMS experiments at the LHC have taken measurements of the production and decays of the Higgs boson and we must ensure that our model is consistent with the existing experimental results. Additionally, we will also consider the impact of projected sensitivities from the high luminosity (HL) LHC. 

To determine these constraints we construct a $\chi^2$ function 
\begin{eqnarray} \label{eq:chi2}
 \chi^2 = \sum_{i,j} \left(\frac{(\sigma\times \text{BR})_i^\text{exp}}{(\sigma\times \text{BR})_i^\text{SM}} - \frac{(\sigma\times \text{BR})_i^\text{BSM}}{(\sigma\times \text{BR})_i^\text{SM}} \right) \left(\frac{(\sigma\times \text{BR})_j^\text{exp}}{(\sigma\times \text{BR})_j^\text{SM}} - \frac{(\sigma\times \text{BR})_j^\text{BSM}}{(\sigma\times \text{BR})_j^\text{SM}} \right) \big( \text{cov} \big)^{-1}_{ij} ~,
\end{eqnarray}
where $(\sigma\times \text{BR})_i^\text{exp}$, $(\sigma\times \text{BR})_i^\text{SM}$, and $(\sigma\times \text{BR})_i^\text{BSM}$ are the experimental measurements, the Standard Model predictions, and our BSM predictions for the production cross sections times branching ratio of the various measured channels.

We use the SM predictions from~\cite{deFlorian:2016spz}. As in \cite{Altmannshofer:2018bch}, the BSM predictions are obtained by rescaling the SM results with appropriate combinations of coupling modifiers. For current LHC results we use the run 1 combination from~\cite{Khachatryan:2016vau}, in addition we take the run 2 results for $h_1\to ZZ^*$~\cite{Sirunyan:2017exp,Aaboud:2017vzb}, $h_1\to WW^*$~\cite{Aaboud:2018jqu}, $h_1\to \gamma\gamma$~\cite{Aaboud:2018xdt,Sirunyan:2018ouh}, $h_1 \to \tau^+ \tau^-$~\cite{Sirunyan:2017khh}, $h_1\to b\bar{b}$~\cite{Sirunyan:2017elk, Aaboud:2017xsd}, $h_1 \to \mu \mu$~\cite{CMS-PAS-HIG-17-019,Aaboud:2017ojs}, and top associated production~\cite{CMS-PAS-HIG-17-003, Aaboud:2017rss}. The projected sensitivities are taken from~\cite{Aaboud:2016cth}, and correspond to $3000$ fb$^{-1}$ of data collected at 14 TeV.

\begin{figure}[tb]
\begin{center} 
\includegraphics[width=0.5\textwidth]{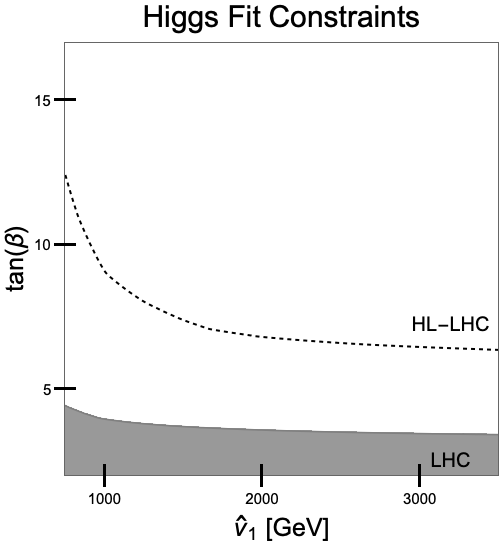}
\caption{The 2$\sigma$ constraints on the twin vev $\hat{v}_1$ vs. $\tan{\beta}$ based on Higgs signal strength measurements at the LHC are shown in the shaded gray region and projections from the HL-LHC are denoted by the black, dotted line. The relevant Higgs potential parameters are set to $\lambda_1 = 1 $, $\lambda_3 = 5$, $\kappa_1 = -3/8$, and $\kappa_2 = 1$. The mass parameters in the flavorful Yukawa couplings are allowed to vary up to a factor of 3 around their expected values.}
\label{HiggsFit}
\end{center}
\end{figure}

The couplings to the SM-like Higgs boson are primarily determined by $\tan{\beta}$ and $\hat{v}_1$. From the expressions in eq.~(\ref{fermionCouplings}) and eq.~(\ref{vectorCouplings}) we can see that generically, large values of $\tan{\beta}$ and $\hat{v}_1$ correspond to couplings of the Higgs to fermions and vector bosons that are SM-like. This can be clearly seen in fig.~\ref{HiggsFit} where we show the Higgs signal strength constraints. We show the $2\sigma$ excluded regions based on the current LHC results and HL-LHC projections. The parameters $\lambda_1$, $\lambda_3$, $\kappa_1$, and $\kappa_2$ that enter the Higgs couplings through the mixing angles in eq.~(\ref{eq:angles}) are set to $\lambda_1 = 1 $, $\lambda_3 = 5$, $\kappa_2 = 1$ with $\kappa_1$ being set to reproduce the SM-like Higgs mass as in eq.~(\ref{higgsMass}). The results are fairly robust to the choice of these parameters, only being modified slightly by the choice of $\lambda_1$ which scales the mixing of the SM-like Higgs with the twin Higgs, as seen in eq.~(\ref{eq:angles}).  In addition, there is also weak dependence of the Higgs couplings on the mass parameters $m_{f_i f_j}$ (see eq.~(\ref{fermionCouplings}) and text below). We let those mass parameters vary up to a factor of 3 around their expected values, as was also done in~\cite{Altmannshofer:2018bch}.

Generally, the sensitivities that are expected at the HL-LHC can potentially constrain the twin vev $\hat{v}_1$ much stronger than the current bound. Previous studies found the constraint $\hat{v}_1 \gtrsim 3 v$, while future experiments favor $\hat{v}_1$ to be closer to an order of magnitude larger than $v$, at least for moderate values of $\tan\beta$. 

As shown in \cite{Altmannshofer:2016zrn}, the flavorful structure we impose on the $\phi_1$ and $\phi_2$ couplings leads to flavor violating Higgs couplings for the SM-like Higgs and the heavy Higgs. The $SU(2)^5$ flavor symmetry, that is preserved by the rank 1 Yukawa couplings of the doublet $\phi_1$, protects flavor changing transitions between the first and second generation of quarks and leptons that typically plague 2HDMs without flavor conservation~\cite{Glashow:1976nt}. However, we still find strong and robust constraints from the rare decay $B_s \to \mu \mu$.\footnote{Other flavor constraints, in particular from $B$ meson oscillations are much less robust, as they depend strongly on the details of the Yukawa couplings of the second Higgs doublet to down type quarks $\lambda_2^d$, see~\cite{Altmannshofer:2018bch}.} In the limit $\hat v_1 \gg v$, the expression for the $B_s \to \mu \mu$ branching ratio in our model can be easily generalized from the expression in~\cite{Altmannshofer:2018bch} with $\alpha \to \alpha_2$. 

The SM prediction and the current experimental measurements are~\cite{Aebischer:2019mlg} 
\begin{eqnarray}
 &&\text{BR}(B_s \to \mu\mu)_\text{SM} = (3.67 \pm 0.15) \times 10^{-9} ~, \nonumber \\
  &&\text{BR}(B_s \to \mu\mu)_\text{exp} = (2.67^{+0.45}_{-0.35}) \times 10^{-9} ~. 
\end{eqnarray}
For the future experimental sensitivities to $B_s \to \mu \mu$, we assume that the central value for the branching ratio stays consistent with the current experimental value, while we take an uncertainty of $\pm 0.16 \times 10^{-9}$~\cite{Albrecht:2017odf}. It is important to note that there is some tension (at the $2\sigma$ level) between the SM prediction and current experimental value, and assuming that the experimental central value holds there will be very significant discrepancy from future experiments. 

\begin{figure}[tb]
\begin{center} 
\includegraphics[width=0.45\textwidth]{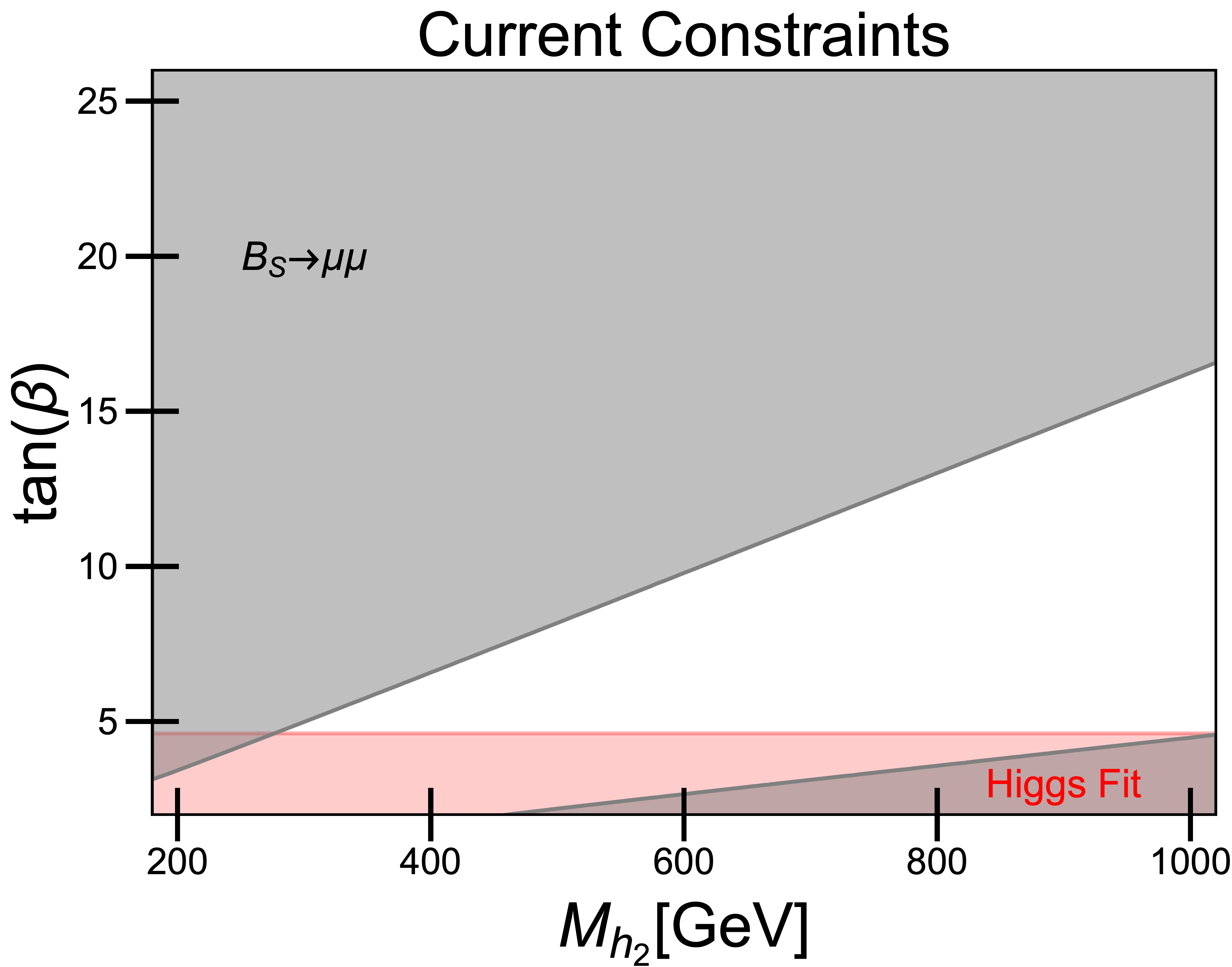} ~~~~~
\includegraphics[width=0.45\textwidth]{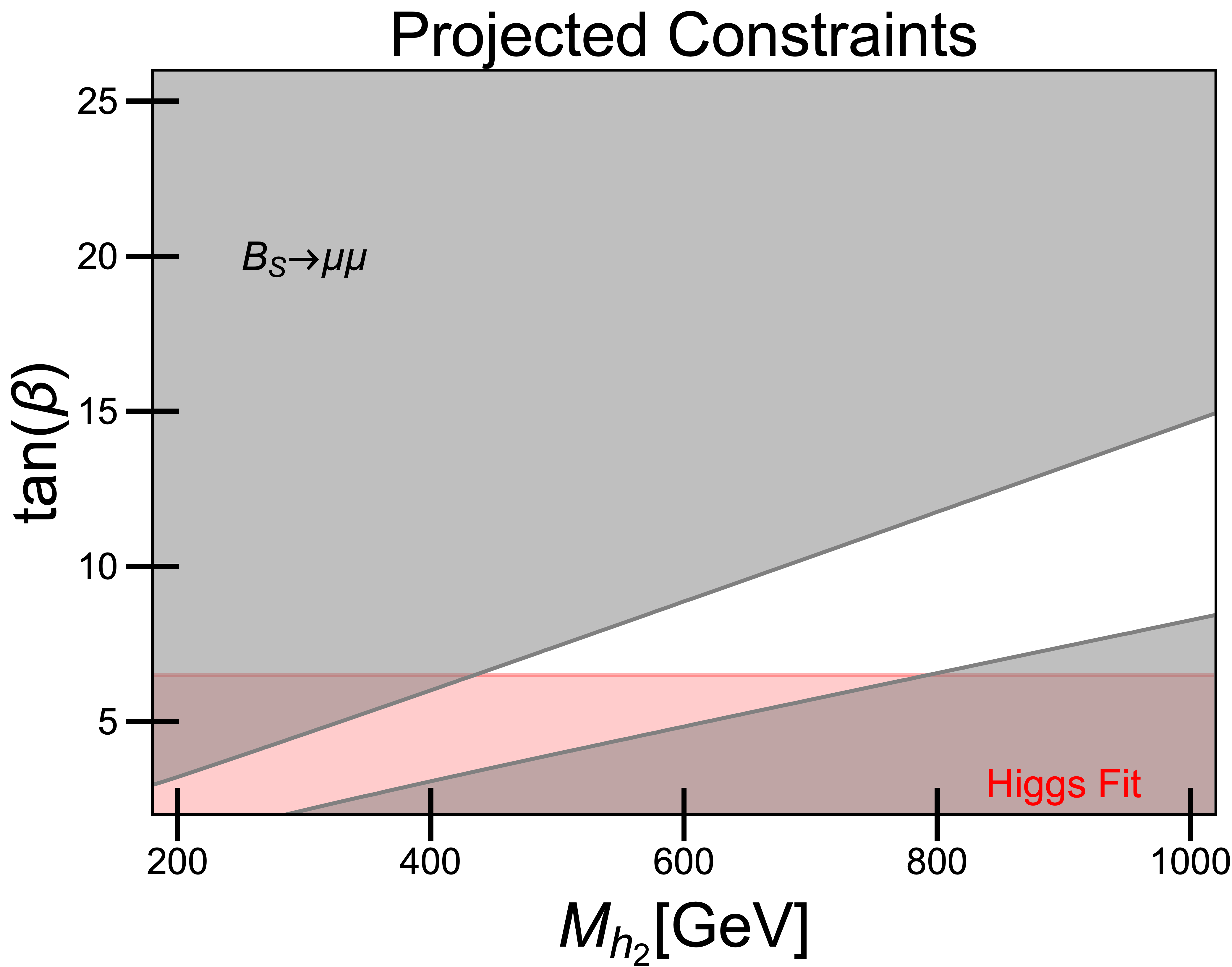}
\caption{Constraints on the heavy Higgs mass $m_{h_2}$ and $\tan{\beta}$ based on current experiments (left) and expected sensitivities (right). We set $\lambda_1 = 1 $, $\lambda_3 = 5$, $\kappa_1 = -3/8$,  $\kappa_2 = 1$, and $\hat{v}_1 = 2500$ GeV.}
\label{HeavyConst}
\end{center}
\end{figure}

The current constraint (left) and future sensitivity (right) from $B_s\to \mu \mu$ is shown in the plane of the heavy Higgs mass $m_{h_2}$ vs. $\tan{\beta}$ in fig.~\ref{HeavyConst}, with the Higgs fit constraints overlayed in red. Based on current constraints masses as low as $300$ GeV are consistent with both $B_s\to \mu \mu$ and Higgs signal strengths measurements for a moderate $\tan{\beta} \simeq 5$. However, future projections push this lower bound on the mass up to around $450$ GeV. For somewhat larger $\tan{\beta} \simeq 10$, the expected bound on the heavy Higgs mass is around $700$ GeV. This is significant as the production of the heavy Higgs becomes quickly suppressed as its mass $m_{h_2}$ increases. 

Our model can moderate the tension between the theoretical prediction and experimental value of the $B_s\to \mu \mu$ branching ratio by the additional contributions of the heavy and pseudoscalar Higgs. In particular the pseudoscalar Higgs contribution interferes destructively with the SM amplitude and thus can lower the $B_s \to \mu \mu$ rate, reconciling the theoretical prediction with the experimental central value. This is evident in the shape of the allowed region of the plots in fig.~\ref{HeavyConst}, where the band represents the region of parameter space that removes unwanted tension. In the scenario that $\tan{\beta}$ becomes too large the rate of $B_s \to \mu \mu$ also becomes too large, violating the $2\sigma$ bound. While if $m_{h_2}$ becomes too large, our theoretical prediction matches back onto the SM prediction and is disfavored.

In addition to Higgs fit and the $B_s \to \mu \mu$ constraints, there exists constraints from heavy Higgs searches performed by ATLAS and CMS. The most relevant constraints come currently from $H \to \mu \mu$ searches~\cite{Aaboud:2017buh, Sirunyan:2018exx}, but are weak compared to the $B_s \to \mu \mu$ and Higgs signal strength measurements as shown in previous work~\cite{Altmannshofer:2018bch}.

\section{Phenomenology}
\label{pheno}

The twin sector in the discussed scenario contains a number of states, the lightest ones being the twin bottom, tau and tau neutrino, as well as twin gluons. The twin gluons and twin bottoms hadronize, with the lightest bound states being bottomonia $[\hat b \bar{\hat b}]$ and glueballs~$\hat G$. For a detailed description of the twin bottomonium and glueball spectrum see~\cite{Craig:2015pha}. 
We will assume that the the twin taus and neutrinos are sufficiently heavy such that the bottomonia and glueballs do not decay into them.
Some of the bottomonia and glueballs (in particular the lightest glueball) can mix with the Higgs bosons in the visible and therefore decay into SM particles. The lifetime of the glueballs can be sizeable and one often finds displaced decays in the discussed scenario.

Displaced events occur as a result of the production of twin sector states (bottomonia and/or glueballs) through one of the three scalar Higgs bosons or the pseudoscalar Higgs boson. We assume that the twin spectrum is such that there are glueball states with mass below half the mass of the twin bottomonia $m_{\hat G} < m_{[\hat b \bar{\hat b}]}/2$ and assume that all decays in the twin sector result in at least one lightest glueball $\hat G_0$. The lightest glueball has the same quantum numbers as the SM-like Higgs allowing it to mix back into the visible sector and decay, in particular to $b\bar b$. In the viable region of our parameter space, the corresponding lifetime of the glueball can be approximated as~\cite{Craig:2015pha}
\begin{equation} \label{eq:lifetime}
c \tau \approx 18m \times \Bigg ( \frac{10 \text{GeV}}{m_{\hat G_0}} \Bigg )^7 \Bigg (   \frac{\hat{v}_1}{750 \text{GeV}}  \Bigg )^4~, 
\end{equation}
which depends very sensitively on the glueball mass.

We can break down the phenomenology of this scenario into three distinct regions: SM-like Higgs dominated, twin Higgs dominated, and heavy Higgs dominated. The SM-like Higgs dominates the phenomenology when the twin vev $\hat{v}_1$ and twin bottom Yukawa $\hat y_{\hat{b}}$ take values such that the twin bottomonia and the twin glueballs are lighter than half the Higgs mass, $m_{[\hat b \bar{\hat b}]} < m_{h_1}/2$. 
As the SM-like Higgs is produced at the LHC at a much higher rate than the heavy Higgs or twin Higgs, displaced decays from the SM-like Higgs dominate the phenomenology\footnote{Note, however, that twin hadrons that are produced from the heavy Higgs or the twin Higgs are much more energetic than those produced from the SM-like Higgs. Therefore, even a small number of displaced decays of the heavy Higgs or twin Higgs might be as prominent as a larger number of displaced SM-like Higgs decays. We do not study such a scenario in detail.}. In this case the phenomenology of our model is similar to that of the original FTH model. For this reason we forgo an analysis of this scenario here and instead point the reader to~\cite{Craig:2015pha, Curtin:2015fna}. 

The twin Higgs dominates the displaced phenomenology when $\hat{v}_1$ and $\hat y_{\hat{b}}$ take on values such that $m_{[\hat b \bar{\hat b}]} >  m_{h_1}/2$, while the twin Higgs $\hat{h}_1$ is still moderately light. In this case the twin Higgs is produced at a high enough rate that its production of twin sector hadrons is much larger than that of the heavy Higgs, so again the phenomenology follows a similar path of the original FTH model, and the addition of the heavy Higgs has little impact on the phenomenology. For this reason we forgo an analysis of this scenario here as well, and instead point the reader to \cite{Craig:2015pha, Ahmed:2017psb, Kilic:2018sew, Alipour-fard:2018mre}. 

The final region of parameter space is characterized by a phenomenology that is dominated by the heavy Higgs. This happens when $\hat{v}_1$ and $\hat y_{\hat{b}}$ take on values such that $m_{[\hat b \bar{\hat b}]}>  m_{h_1}/2$ and at the same time the twin Higgs $\hat{h}_1$ is very heavy (motivated by $\hat{v}_1$ being large). In this regime both the heavy Higgs and the twin Higgs participate in producing twin sector particles, but the production rate of the twin Higgs becomes very small. As $\hat{v}_1$ rises, the decay of the heavy Higgs to the twin sector is also suppressed. However, the production rate of the twin Higgs drops more quickly than the branching ratio of heavy Higgs decays into twin sector particles. The result of this is a region of parameter space where the heavy Higgs plays the most important role in the phenomenology. 

\begin{figure}[tb]
\begin{center} 
\includegraphics[width=0.47\textwidth]{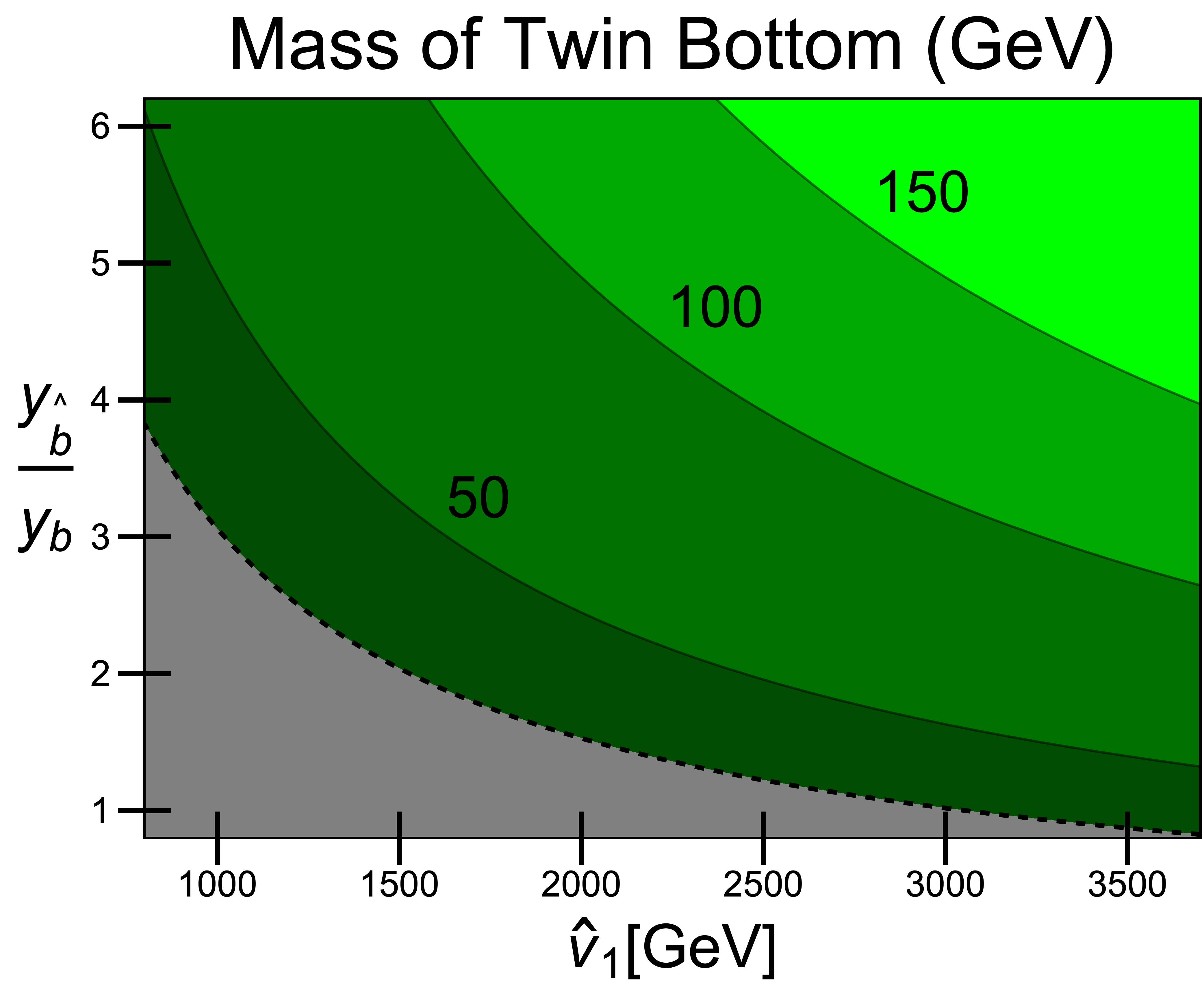}
\caption{The mass of the twin bottom as a function of the twin bottom Yukawa coupling $y_{\hat{b}}$ and twin vev $\hat{v}_1$. In the gray region twin bottoms are sufficiently light such that the SM-like Higgs can decay into twin bottomonia.}
\label{twinBottomMass}
\end{center}
\end{figure}

The mass of the twin bottom as a function of the coupling twin bottom Yukawa coupling $y_{\hat{b}}$ and the twin vev $\hat{v}_1$ is shown in fig.~\ref{twinBottomMass}. The gray region denotes where the twin bottom is light enough, such that bottomonium can be produced by the SM-like Higgs. The phenomenology of this region is analogous to that of traditional twin Higgs models. In the following we focus on the region with heavy twin bottoms $m_{\hat{b}} \sim O(100\,\text{GeV})$.

\begin{figure}[tb]
\begin{center} 
\includegraphics[width=0.47\textwidth]{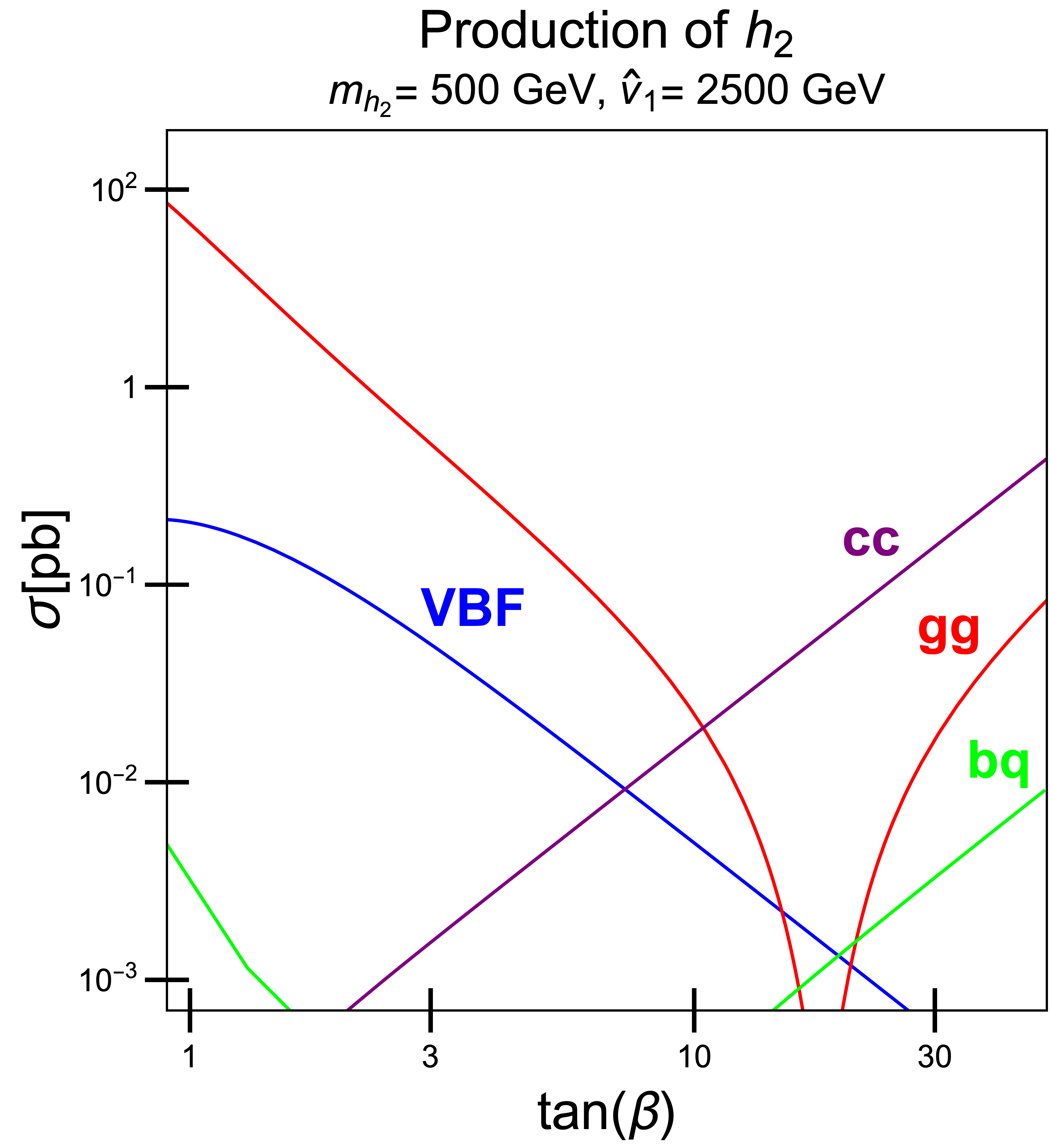}
\includegraphics[width=0.47\textwidth]{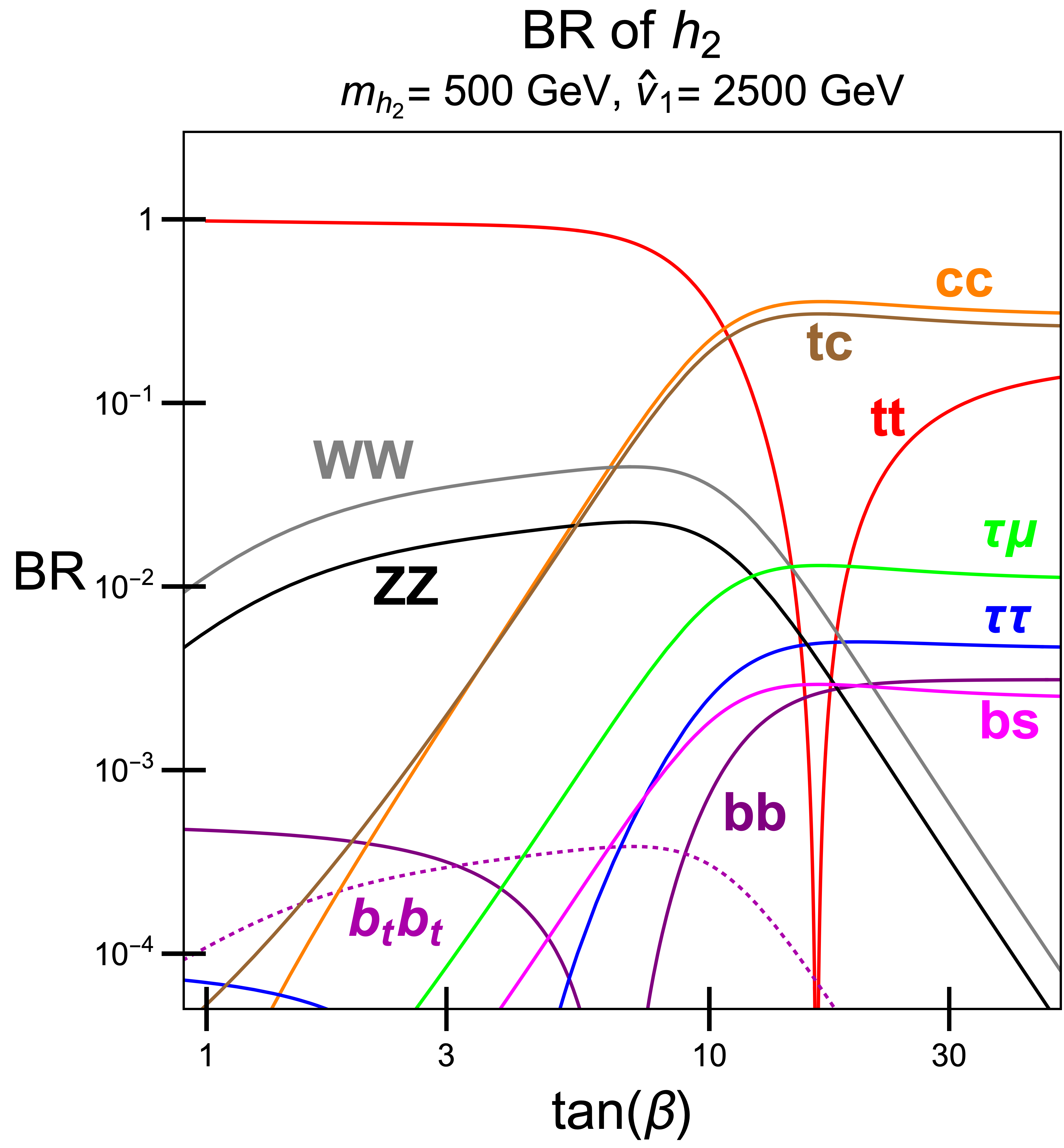}
\caption{The production cross section at 13 TeV proton-proton collisions (left) and branching ratios (right) of the heavy Higgs with mass $m_{h_2} = 500$ GeV as function of $\tan\beta$. We set $\kappa_1 = -3/8, \kappa_2 = 1, \lambda_1 = 1$, $\lambda_3 = 5$, $\hat{v}_1 = 2500$ GeV, and $\hat y_{\hat b} = 3 y_b^\text{SM}$.}
\label{heavyPheno}
\end{center}
\end{figure}

To better understand the production of the twin bottom via the heavy Higgs boson we look at the production and decay of the heavy Higgs boson in fig.~\ref{heavyPheno}. The right plot shows the branching ratios of the heavy Higgs with mass $m_{h_2} = 500$ GeV as a function of $\tan{\beta}$, for a benchmark parameter point defined by $\kappa_1 = -3/8, \kappa_2 = 1, \lambda_1 = 1$, $\lambda_3 = 5$, and $\hat{v}_1 = 2500$ GeV. The production cross sections and decay rates are rather robust against order one changes to the parameters of this benchmark point other than $\lambda_3$. We can see from eq.~(\ref{eq:angles}) that $\lambda_3$ controls the mixing of the heavy Higgs with the twin sector and therefore has a substantial impact on the branching ratio into the twin sector BR$(H\to \hat{b} \hat{b})$. We choose a large value of $\lambda_3 = 5$ as a representative example. For the Yukawa coupling of the twin bottom we choose $\hat y_{\hat b} = 3 y_b^\text{SM}$, within the range allowed by naturalness arguments, see eq.~(\ref{yfConst}). The branching ratios in this model are similar to that of the type 1B F2HDM~\cite{Altmannshofer:2018bch}, with the addition of a small, but important, branching ratio to the twin bottom. In the left plot we show the production cross sections of the heavy Higgs. Also the cross sections are similarly to the type 1B F2HDM and we see that over most of parameter space the main production modes are charm-charm fusion, gluon-gluon fusion, and vector boson fusion. Generally, gluon-gluon fusion is dominant at small $\tan{\beta}$ and charm-charm fusion is dominant at high $\tan{\beta}$. 
For moderate values of $\tan\beta$, the production cross section for a $500$~GeV heavy Higgs can be around $100$~fb.

\begin{figure}[tb]
\begin{center} 
\includegraphics[width=0.9\textwidth]{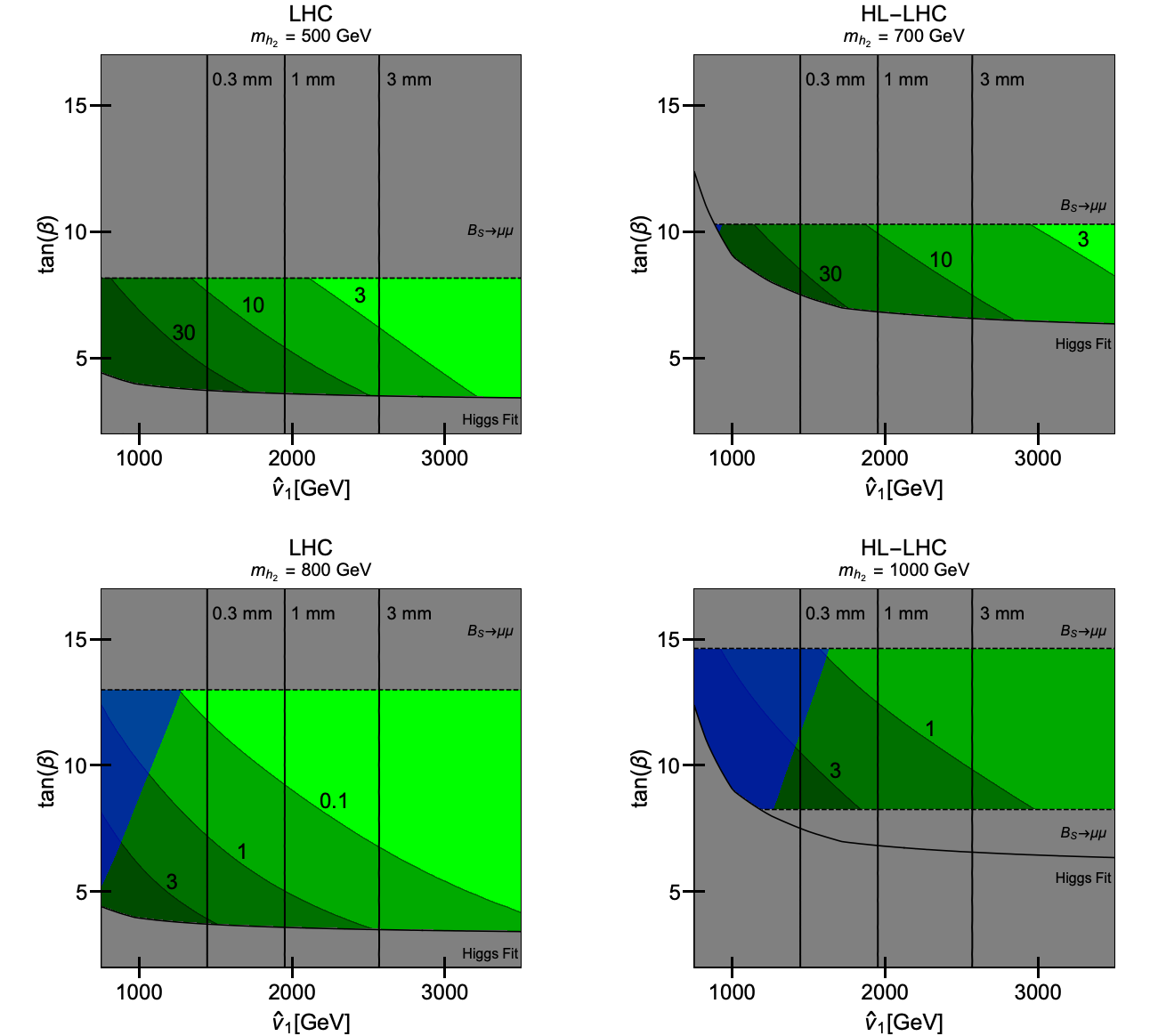}
\caption{The allowed parameter space for the process $pp \to H \to \hat{b}\bar{\hat{b}}$ in the plane of twin Higgs vev $\hat v_1$ vs. $\tan\beta$. The left two plots show the current constraints as well as the prediction for the number of events (green contours) after run 3 of the LHC for a heavy Higgs mass of 500~GeV (top) and 800~GeV (bottom). The right two plots show the expected constraints and predictions of number of events at the HL-LHC with 3000\,fb$^{-1}$ for a heavy Higgs mass of 700~GeV (top) and 1000~GeV (bottom). The gray shaded region with a solid black boundary shows the exclusion due to Higgs signal strength fits and the gray shaded region with the dashed black boundary shows the constraints from the $B_s \to \mu \mu$ decay. In the blue shaded region the number of displaced decays coming from the twin Higgs exceeds that of the heavy Higgs. The vertical black contours show the proper lifetime of twin glueballs with mass $70$\,GeV.}
\label{twinSectorDecays}
\end{center}
\end{figure}

The number of expected events with displaced decays as a function of the twin vev $\hat{v}_1$ and $\tan{\beta}$ is shown in fig.~\ref{twinSectorDecays}. We assume that each decay of the heavy Higgs into twin bottoms results in at least one long lived glueball that decays back into the SM through mixing with a Higgs. The left two plots show the number of events produced after run 3 of the LHC (300/fb) for a heavy Higgs mass of 500~GeV (top) and 800~GeV (bottom), while the right two plots shows the number produced displaced decay events after the conclusion of the HL-LHC (3000/fb) for a heavy Higgs mass of 700~GeV (top) and 1000~GeV (bottom). The chosen heavy Higgs masses correspond to choices which still exhibit a fair amount of freedom in other parameters, such as $\tan{\beta}$ as shown in fig.~\ref{HeavyConst}. The gray shaded region corresponds to the parameter space ruled out by either the $B_s \to \mu \mu$ constraint (dashed boundary), or Higgs signal strength measurements (solid boundary). The blue region shows where the phenomenology is dominated by the twin Higgs. 

The vertical black lines show the proper lifetime $c\tau$ of the twin glueballs. The lifetime is primarily determined by $\hat{v}_1$ and the mass of the glueballs, see eq.~(\ref{eq:lifetime}). In fig~\ref{twinSectorDecays} we set the glueball mass to $70$\,GeV, such that decays of the SM-like Higgs to glueballs are completely absent. In that case, when $\hat{v}_1$ is larger than about 2000 GeV, we see that the lifetime of the glueballs is of the order of at least millimeters which, given a typical boost factor of a few, falls into the decay lengths of interest for displaced signatures at the LHC. 

We see that for the lower mass choices of the heavy Higgs $500$ GeV (at the LHC) and $700$ GeV (at the HL-LHC) $O(10)$s of events could occur with $O(\text{few~mm})$ displaced decays. As we push the scale of $\hat{v}_1$ to larger values, we see that this number drops down to a handful of decays. The HL-LHC will generically produce more displaced decays at a given heavy Higgs mass, but the stronger expected constraints on the parameter space roughly balance out the increase. So, for masses that are not indirectly probed by flavor constraints or Higgs coupling strength measurements, we see a similar amount of expected displaced decays. Similarly such an observation also holds for the higher mass scenarios that we considered. We see that for heavier Higgs mass $m_{h_2}$ the estimated number of events is reduced to several and below as the production of the heavy Higgs is suppressed at these higher masses. 

Searches for long lived particles have been explored to some degree at the LHC \cite{Lee:2018pa, Alimena:2019zri}.
The existing searches do currently not put strong constraints on the displaced decays we have considered in this model. For sizable displacements of the order $O(1\,\text{cm} - 1\,\text{m})$ the expected sensitivities from the LHC could cover sizable regions of parameter space (see~\cite{Alipour-fard:2018mre} for a detailed study of the fraternal twin Higgs model). However, in the scenarios discussed in this paper, the displacement is typically of the order $O(\text{few}~\text{mm})$\footnote{Note however that the displacement depends strongly on the assumed glueball mass, see eq.~(\ref{eq:lifetime}). Reducing the glueball mass slightly, say to 60 GeV, increases the lifetime of the glueballs by a factor of $(7/6)^7 \simeq 3$.}, making it much more challenging to search for the displaced signatures, for example due to triggering difficulties (see e.g.~\cite{Craig:2015pha, Alimena:2019zri}). Future improvements in searching for displaced decays with $O(\text{few}~\text{mm})$ displacement would be necessary to further explore the models described in this work.

\section{Conclusion} \label{sec:conclusions}

The little hierarchy problem and the SM flavor puzzle are two longstanding problems in particle physics.
We have discussed a setup which attempts to address both of them (at least partially).
We considered a 2HDM with a flavorful Yukawa structure, where one Higgs doublet is responsible for the mass of the third generation fermions and the other doublet is responsible for the mass of the first and second generations. A hierarchy in vevs can explain the mass hierarchy between the third and first two generations. We combined this setup with the twin Higgs mechanism which stabilizes the Higgs mass up to $O(10\,\text{TeV})$, considering both a mirror twin and fraternal twin setup. 

In the visible sector, the flavorful Yukawa structure of this model leads to modifications of the $B_s \to \mu \mu$ branching ratio. Large values of $\tan\beta$ are already strongly constrained. We showed that the current mild tension that exists between the SM prediction and the experimental results can be solved in our setup for moderate values of $\tan\beta$. This is of particular interest in view of the expected future sensitivities to $B_s \to \mu \mu$ from LHCb, which could turn the current tension into a very significant discrepancy.

The second (heavy) Higgs doublet in the visible sector also provides interesting phenomenology in the form of displaced signatures at the LHC.
This heavy Higgs can decay into the twin sector, in particular twin bottomonia and twin glueballs (which we assume to be the lightest states in the twin sector in this work) that can subsequently decay back to the visible sector through mixing with the Higgs bosons. This often leads to displaced signatures, in particular displaced $b$-jets.
This can happen in regions of parameter space in which displaced signatures of the SM-like Higgs and the twin Higgs are suppressed or even completely absent. The corresponding parameter space is characterized by a heavy twin sector, where the production cross section of the twin Higgs is small and the SM-like Higgs is kinematically excluded from decaying to the twin sector. We have shown that in such a scenario the heavy Higgs boson can be light enough to be produced with a sizeable cross section and heavy enough to decay into the twin sector, thus offering the possibility to probe broader regions of parameter space with searches for displaced signatures. 

The prediction of this scenario are slightly displaced decays at length scales of few millimeters which are challenging to detect experimentally. We find that for a twin vev $\hat{v}_1$ of at least 2000 GeV that the heavy Higgs can naturally dominate the displaced phenomenology with as many as $O(30)$ displaced decays predicted to have taken place at the LHC already. Anticipating improved indirect constraints on the model parameter space from future experimental results on Higgs signal strengths measurements and the $B_s \to \mu \mu$ decay, we find that there is still viable region of parameter space which can produce $O(30)$ displaced decays at the HL-LHC.

\section*{Acknowledgements}

We would like to thank Stefania Gori for useful discussions.
The research of WA and BM is supported by the National Science Foundation under Grant No. PHY-1912719.

\bibliographystyle{ieeetr} 
\bibliography{refs}

\end{document}